\pdfoutput=1

\documentclass[10pt,twocolumn]{article}

\usepackage[numbers,sort&compress]{natbib}
\usepackage[labelfont=bf,labelsep=period]{caption}
\usepackage{hyperref}
\usepackage{graphicx}
\usepackage{varwidth}
\usepackage{subfigure}
\usepackage{booktabs}
\usepackage{amssymb}
\usepackage{multicol}
\usepackage{xspace}
\usepackage{url}
\usepackage{paralist}
\usepackage{todonotes}
\usepackage{authblk}

\graphicspath{{./pictures/}}

\usepackage[T1]{fontenc}
\usepackage[tt=false]{libertine}
\usepackage[libertine]{newtxmath}
\usepackage{inconsolata}
\usepackage{geometry}

\hypersetup{ colorlinks = true, citecolor = red,
  urlcolor = cyan, linkcolor = red
}

\usepackage{booktabs}

\title{\textbf{LibrettOS: A Dynamically Adaptable Multiserver-Library OS}\footnote{\textcopyright 2020 Copyright held by the owner/author(s). Publication rights licensed to ACM. This is the author's version of the work. It is posted here for your personal use. Not for redistribution. The definitive Version of Record was published in Proceedings of the 16th ACM SIGPLAN/SIGOPS International Conference on Virtual Execution Environments (VEE '20), March 17, 2020, Lausanne, Switzerland \url{http://dx.doi.org/10.1145/3381052.3381316}.\newline\newline The U.S. Government is authorized to reproduce and distribute reprints for Governmental purposes notwithstanding any copyright annotation thereon.
}}

\author{Ruslan Nikolaev, Mincheol Sung, Binoy Ravindran}
\affil{Bradley Department of Electrical and Computer Engineering, Virginia Tech\\ $\{ rnikola, mincheol, binoy \} $@vt.edu}

\date{}

\begin{document}

\maketitle

\begin{abstract}
We present LibrettOS, an OS design that fuses two paradigms
to simultaneously address issues of isolation, performance, compatibility, failure recoverability,
and run-time upgrades. LibrettOS acts as a
microkernel OS that runs servers in an isolated manner.
LibrettOS can also act as a library OS when,
for better performance, selected applications
are granted exclusive access to virtual hardware resources such
as storage and networking.
Furthermore, applications can switch between the two OS modes
with no interruption at run-time.
LibrettOS has a uniquely distinguishing advantage in that, the two paradigms seamlessly coexist in the same OS, enabling users to simultaneously exploit their respective strengths (i.e., greater isolation, high performance).
Systems code, such as device drivers,
network stacks, and file systems remain  identical in the two modes, enabling dynamic mode switching and reducing development and maintenance costs.

To illustrate these design principles, we implemented a prototype of LibrettOS using rump kernels, allowing us to reuse existent, hardened NetBSD device drivers
and a large ecosystem of POSIX/BSD-compatible applications.
We use hardware (VM) virtualization to strongly isolate different
rump kernel instances from each other.
Because the original rumprun unikernel targeted a much simpler model for uniprocessor systems,
we redesigned it to support multicore systems.
Unlike kernel-bypass libraries such as DPDK, applications need not be modified to benefit from direct hardware access. LibrettOS also supports indirect access through a network server that we have developed. Instances of the TCP/IP stack always run directly inside the address space of applications. Unlike the original rumprun or monolithic OSes, applications remain uninterrupted even when network components fail or need to be upgraded. Finally, to efficiently use hardware
resources, applications can
dynamically switch between the  indirect and direct modes  based on their I/O load at  run-time.
We evaluate LibrettOS with 10GbE and NVMe using Nginx, NFS, memcached, Redis, and other applications. LibrettOS's performance
typically exceeds that of NetBSD, especially when using direct access.

\end{abstract}

\def\keywords{\vspace{.5em}
{\noindent{\textbf{\textit{Keywords}:}\,\relax%
}}}
\def\endkeywords{\par}

\keywords{operating system, microkernel, multiserver, network server, virtualization, Xen, IOMMU, SR-IOV, isolation}

\newcommand{\netdom}{
\begin{figure}[ht!]
\centering
\includegraphics[width=.9\columnwidth]{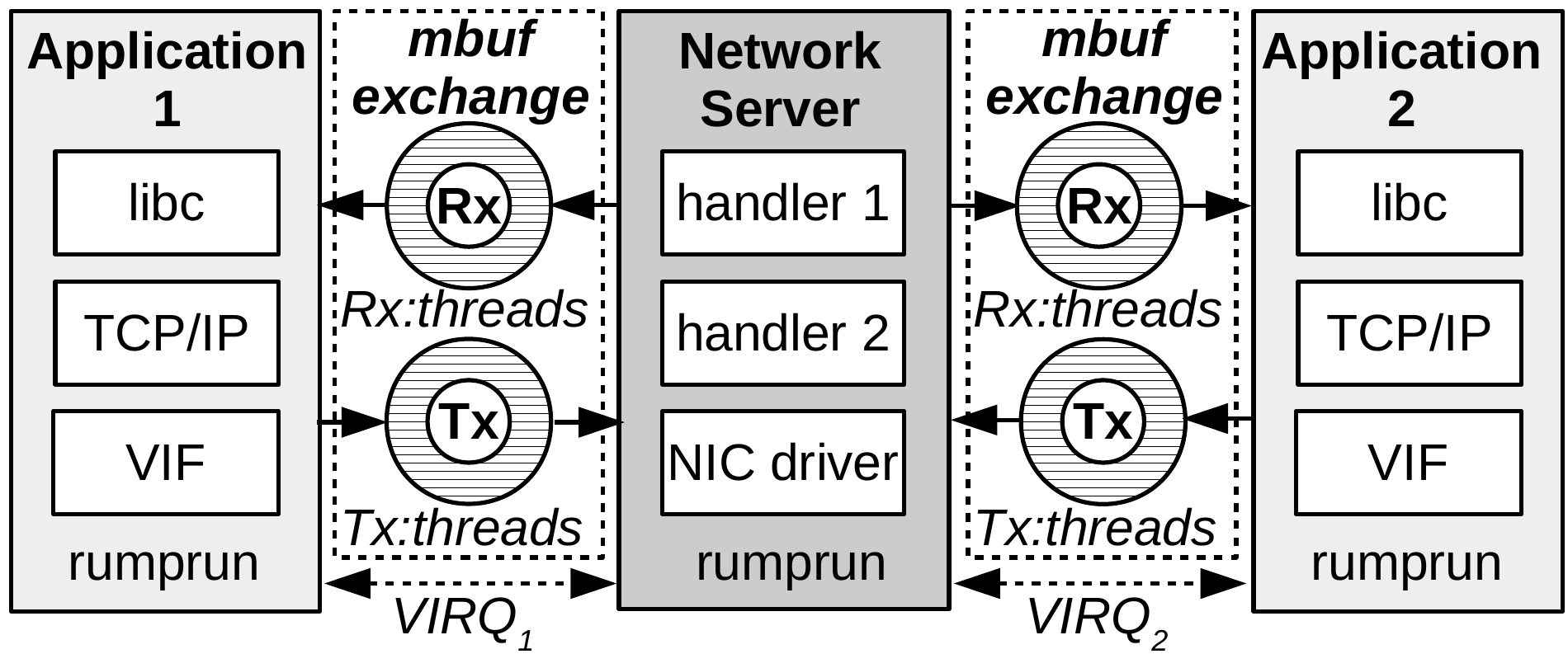}
\caption{Network server.}
\label{fig:netdom}
\end{figure}
}

\newcommand{\app}{
\begin{figure}[ht!]
\centering
\includegraphics[width=.9\columnwidth]{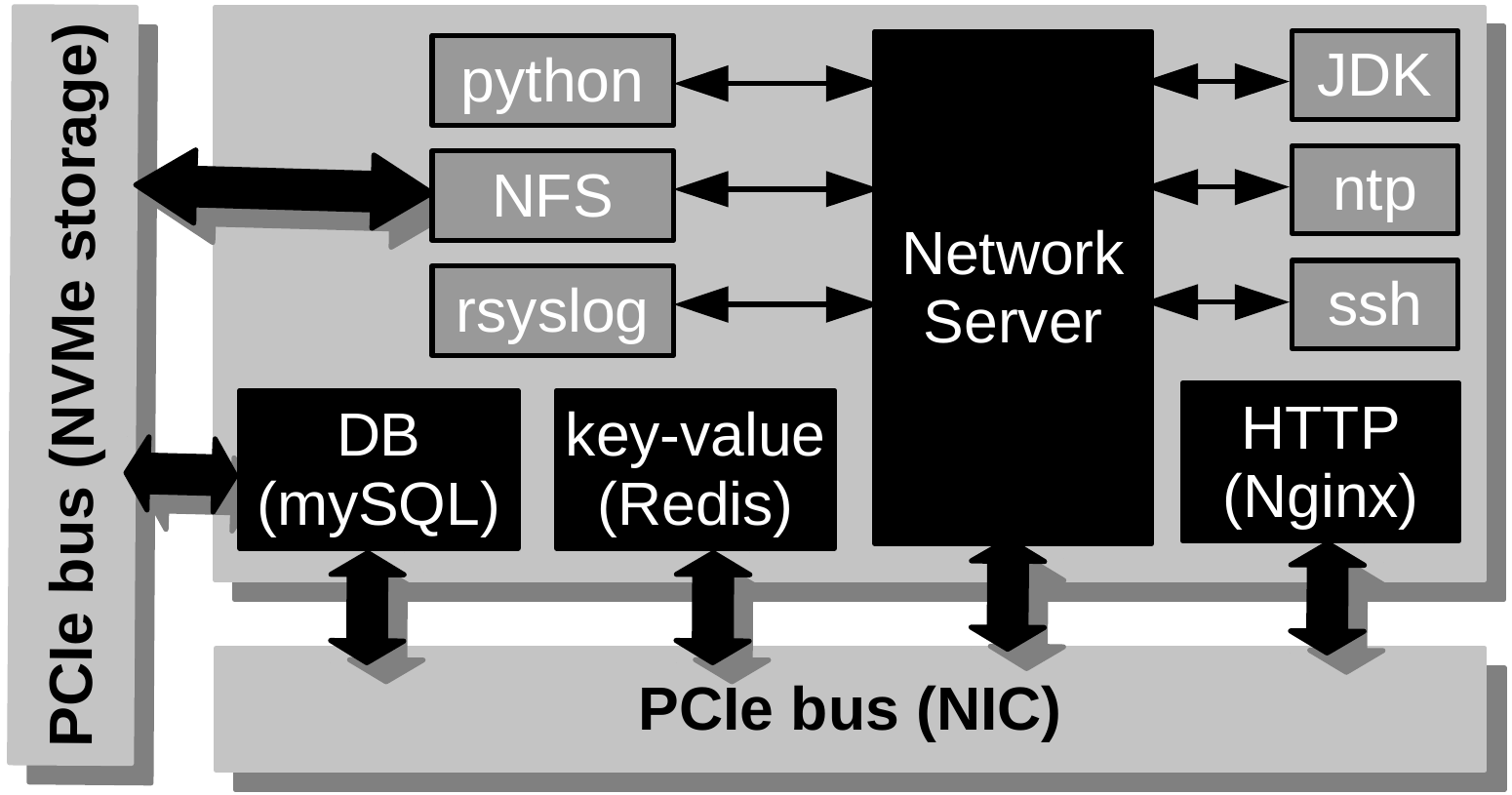}
\caption{Server ecosystem example.}
\label{fig:app}
\end{figure}
}

\newcommand{\rump}{
\begin{figure}[ht!]
\centering
\includegraphics[width=.8\columnwidth]{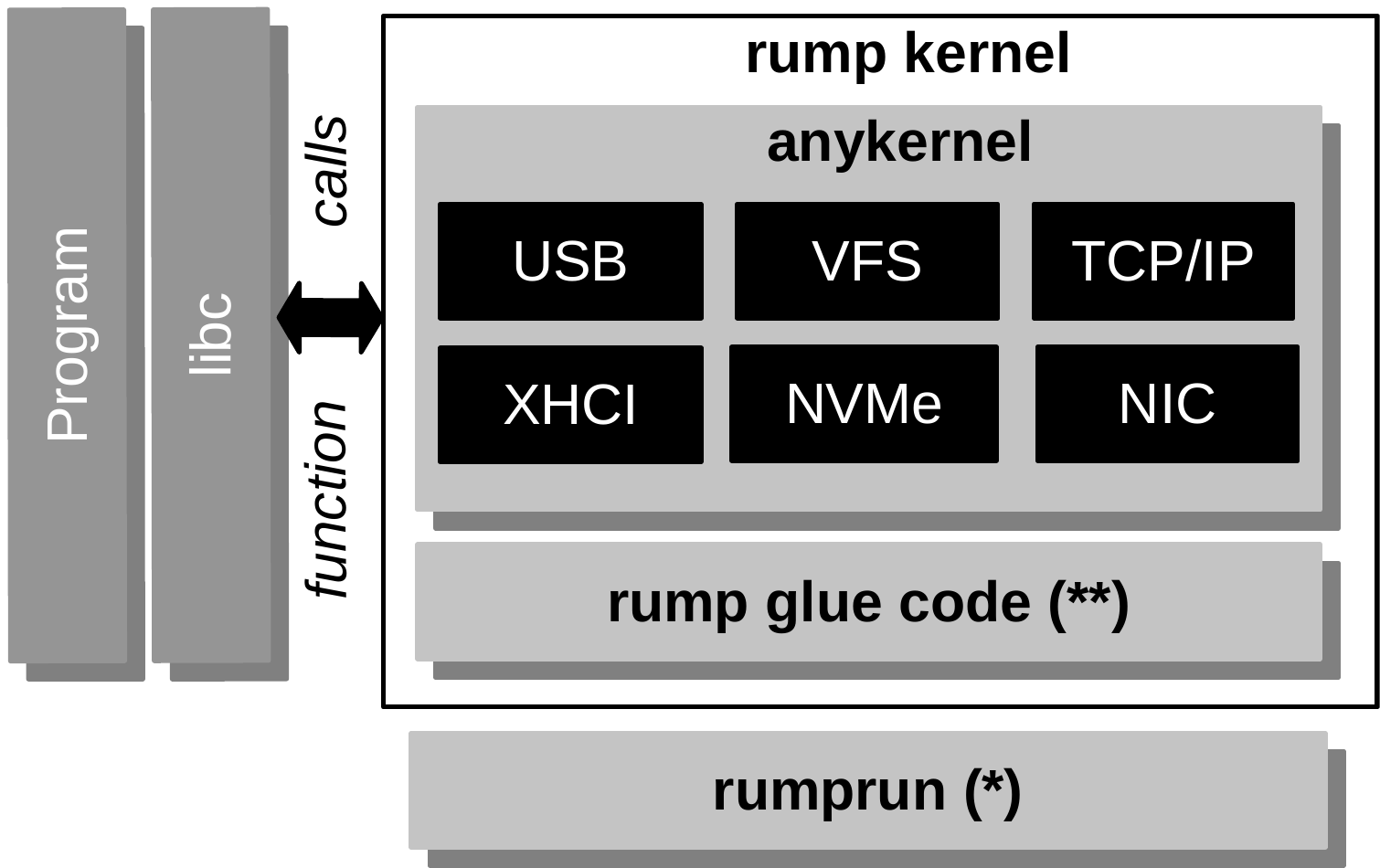}
\caption{Rump software stack, (*) indicates components substantially and (**) partially modified in LibrettOS.}
\label{fig:rump}
\end{figure}
}
\newcommand{\librettos}{
\begin{figure}[ht!]
\centering
\includegraphics[width=.9\columnwidth]{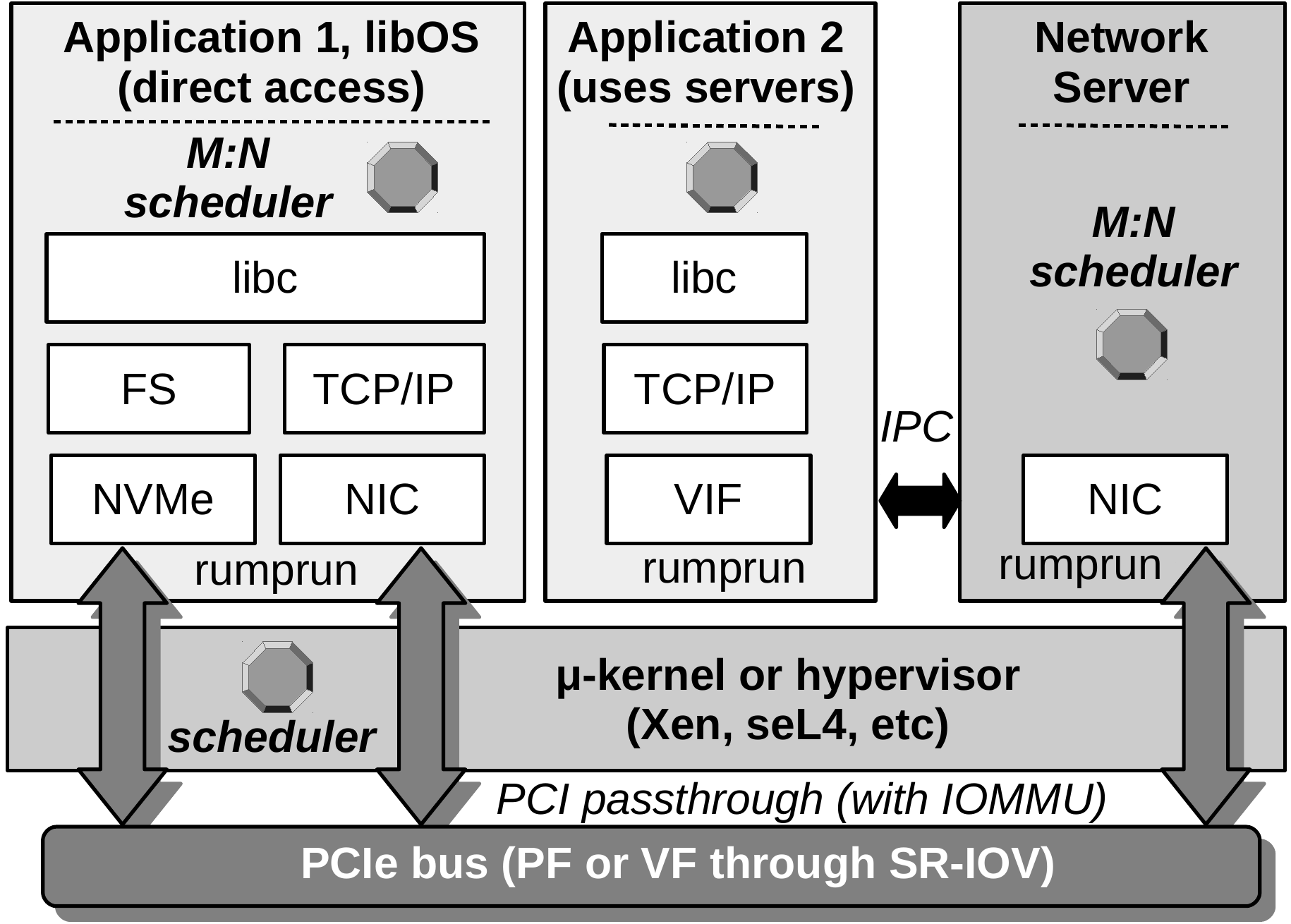}
\caption{LibrettOS's architecture.}
\label{fig:librettos}
\end{figure}
}

\section{Introduction}

Core components and drivers of a general purpose monolithic operating system (OS)
such as Linux or NetBSD typically run in privileged mode. However, this design
is often inadequate for modern systems~\cite{MINIX,HELENOS,VIRTUOS,RUMP,HERMITCORE}. On the one hand, a
diverse and ever growing kernel ecosystem requires better isolation of individual drivers
and other system components to localize security threats due to the increasingly
large attack surface of OS kernel code. Better isolation also helps with
tolerating component failures and thereby increases  reliability. Microkernels achieve this goal, specifically in multiserver 
OS designs~\cite{MINIX,QNX,HELENOS}.\footnote{As microkernels are defined broadly in the literature, we
clarify that we consider multiserver OSes as those implementing \textit{servers} to isolate core OS components, e.g., MINIX 3~\cite{MINIX}.}
On the other hand, to achieve better device throughput and resource utilization, some
applications need to bypass the system call and other layers so that they can
obtain exclusive access to device resources such as network adapter's (NIC) Tx/Rx queues.
This is particularly useful in recent hardware with SR-IOV support~\cite{SRIOV},
which can create virtual PCIe functions: NICs or NVMe storage
partitions.
Library OSes and kernel-bypass libraries~\cite{DPDK, SPDK}  achieve this goal.
Multiserver-inspired designs, too, can outperform
traditional OSes on recent hardware~\cite{SNAP,FSPROCESSES}.

\app{}

Microkernels, though initially slow in adoption, have gai\-ned more traction
in recent years. 
Google's Fuchsia OS~\cite{FUCHSIA} uses the Zircon microkernel. Intel Management Engine~\cite{INTELME} uses MINIX 3~\cite{MINIX}
since 2015.
A multiserver-like networking design was recently revisited
by Google~\cite{SNAP} to improve performance and upgradability when using their private
(non-TCP) messaging protocol.
However, general-purpose application and device driver support for microkernels is limited,
especially for high-end hardware such as 10GbE+.

Kernel-bypass techniques, e.g., DPDK~\cite{DPDK} and SPDK~\cite{SPDK}, are also increasingly gaining traction, as they
eliminate OS kernels from the critical data path, thereby improving
performance.
Unfortunately, these techniques lack standardized high-level APIs
and require massive engineering effort to use, especially to adapt to existing 
applications~\cite{Zhang:2019:IDY:3317550.3321422}.
Additionally, driver support in kernel-bypass libraries such as DPDK~\cite{DPDK} is great only for high-end NICs from certain vendors.
Re-implementing drivers and high-level OS stacks
from scratch in user space involves significant  development effort.

Oftentimes, it is overlooked that ``no one size fits all.'' In other words, no single OS model is ideal for all use cases.  
Depending upon the application, security  
or reliability requirements, it is desirable to employ multiple OS paradigms in
the \textit{same} OS. In addition, applications
may need to switch between different OS modes based on their I/O loads.
In Figure~\ref{fig:app}, we illustrate 
an ecosystem of a web-driven server running on the same physical
or virtual host. The server uses tools for logging (rsyslog), clock synchronization (ntp), NFS shares, and SSH for remote access. The server
also runs python and Java applications. None of these
applications are performance-critical, but due to the complexity of the network and
other stacks, it is important to recover from temporary failures or
bugs without rebooting, which is impossible 
in monolithic OSes. One way to meet this goal is to have a network server as
in the multiserver paradigm, which runs system components
in separate user processes for better isolation and failure recoverability.
This approach is also convenient when
network components need to be upgraded and restarted at run-time~\cite{SNAP} by triggering an artificial fault.

Core applications such as an HTTP server, database, 
and key-value store are more I/O performance-critical.
When having a NIC and NVMe storage with SR-IOV support (or multiple devices),
selected applications can access hardware resources directly
as in library OSes or kernel-bypass libraries -- i.e., by bypassing the system call
layer or inter-process communication (IPC). 
Due to finite resources, 
PCIe devices
restrict the number of SR-IOV interfaces -- e.g., the Intel 82599 adapter~\cite{INTEL82599}
supports up to 16 virtual NICs, 4 Tx/Rx queues each; for other adapters, 
this number can be even smaller.
Thus, it is important to manage available hardware I/O resources efficiently.
Since I/O load is usually non-constant and changes for each application based
on external circumstances (e.g., the number of clients connected to an HTTP server during peak and normal
hours), conservative resource management is often desired: use network server(s)
until I/O load increases substantially, at which point, migrate to the library OS mode (for direct access)  at run-time with no interruption.
This is especially useful for recent bare metal cloud systems -- e.g., when one
Amazon EC2 bare metal instance is shared by several users.

In this paper, we present a new OS design -- LibrettOS -- that not only reconciles the
library and multiserver OS paradigms while retaining their
individual benefits (of better isolation, failure recoverability, and performance),
but also overcomes their downsides (of driver and application incompatibility).
Moreover, LibrettOS enables
applications to switch between these two
paradigms at run-time.
While high performance can be obtained with specialized APIs, which can also be adopted in LibrettOS, they
incur high engineering effort.
In contrast, with LibrettOS, existing applications can already benefit from more direct access to hardware while still using POSIX.

We present a realization of the LibrettOS design through a prototype implementation. Our prototype leverages rump kernels~\cite{Kantee_flexibleoperating}
and reuses a fairly large NetBSD driver collection. Since
the user space ecosystem is also inherited from NetBSD, the prototype retains excellent 
compatibility with existing POSIX and BSD applications as well.
Moreover, in the two modes of operation (i.e., library OS mode and multiserver OS mode), we use an identical set of
drivers and software.
In our prototype, we focus only on networking and storage. However,
the LibrettOS design principles are more general, as rump kernels can potentially
support other subsystems -- e.g., NetBSD's sound drivers~\cite{HURDSOUND} can be reused.
The prototype builds on rumprun instances, which execute
rump kernels atop a hypervisor.
Since the original rumprun did not support multiple cores,
we redesigned it to support symmetric multiprocessing (SMP)
systems. We also added 10GbE and NVMe drivers, and made other
improvements to rumprun.
As we show
in Section~\ref{exp:evalsect}, the prototype outperforms the original
rumprun and NetBSD, especially
when employing direct
hardware access. In some tests, the prototype also outperforms Linux, which is often better optimized for performance than NetBSD.

The paper's \textbf{research contribution} is the proposed OS design
and its prototype. Specifically, LibrettOS is the first OS  design that \textit{simultaneously}  supports the
multiserver and library OS paradigms; it is also the first design that can \textit{dynamically switch}  between these two OS modes at run-time with no interruption to applications.
Our network server, designed for the multiserver mode,
accommodates both paradigms by using
fast L2 frame forwarding from applications. Finally, LibrettOS's design requires only \textit{one set
of device drivers} for both modes and mostly retains POSIX/BSD compatibility.

Our work also confirms an existent hypothesis that kernel bypass is
faster than a monolithic OS approach. However, unlike prior works~\cite{IX,SNAP,ARRAKIS},
we show this without resorting to optimized (non-POSIX) APIs or specialized device drivers.

\section{Background}

For greater clarity and completeness, in this section, we discuss various OS designs, microkernels, and hypervisors. We also provide background information on rump kernels and their practical use for hypervisors and bare metal machines.

\begin{table*}
\begin{center}
	\begin{tabular}{ l l l l l l l l l }
    \toprule
		\textbf{System} & \textbf{API} & \textbf{Generality} & \textbf{TCP} & \textbf{Driver} & \textbf{Paradigms} & \textbf{Dynamic} & \textbf{Direct} & \textbf{Failure} \\
 & & & \textbf{Stack} & \textbf{Base} & & \textbf{Switch} & \textbf{Access} & \textbf{Recovery} \\
    \midrule
        DPDK~\cite{DPDK} & Low-level & Network & 3rd party & Medium & N/A & No & Yes & No \\
        SPDK~\cite{SPDK} & Low-level & Storage & N/A & NVMe & N/A & No & Yes & No \\
        IX~\cite{IX} & Non-standard & Network & App & DPDK & Library OS & No & Almost & No \\
        Arrakis~\cite{ARRAKIS} & POSIX/Custom & Network & App & Limited & Library OS & No & Yes & No \\
	Snap~\cite{SNAP} & Non-standard & Network & unavailable & Limited & Multiserver & No & No & Yes \\
	VirtuOS~\cite{VIRTUOS} & POSIX/Linux & Universal & Server & Large & Multiserver & No & No & Yes \\
        MINIX 3~\cite{MINIX} & POSIX & Universal & Server & Limited & Multiserver & No & No & Yes \\
        HelenOS~\cite{HELENOS} & POSIX & Universal & Servers & Limited & Multiserver & No & No & Yes \\
        Linux & POSIX/Linux & Universal & Kernel & Large & Monolithic & No & No & No \\
        NetBSD~\cite{NETBSD} & POSIX/BSD & Universal & Kernel & Large & Monolithic & No & No & No \\
        \textbf{LibrettOS} & POSIX/BSD & Universal & App & Large & Multiserver, & \textbf{Yes} & \textbf{Yes} & \textbf{Yes} \\
         &  &  & & (NetBSD) & Library OS &  &  \\
	\bottomrule
\end{tabular}
\end{center}
\caption{Comparison of LibrettOS with libraries and frameworks as well as monolithic, multiserver, and library OSes.}
\label{tbl:comparos}
\end{table*}

\subsection{Multiserver and Library OSes}

Traditionally, monolithic OSes run all critical components in a privileged CPU mode within a single kernel address space that is isolated from user applications running in an unprivileged mode.
The isolation of systems software components such as device drivers
in separate address spaces is the fundamental principle behind the microkernel model~\cite{NUCLEUS, MINIX, L4,
MACH}, which provides stronger security and reliability~\cite{SEL4,
MINIX_FAULT_ISOLATION, MINIX_ASLR} guarantees to applications compared to the
classical monolithic kernel designs. To support existing applications and drivers, microkernel systems either implement emulation
layers~\cite{Hand+:hotos2005} or run a multiserver
OS~\cite{Gefflaut+:eurosigops2000,Herder+:asci2006,MINIX,QNX,HELENOS} on top of the microkernel.
In the multiserver design, servers are typically created for specific
parts of the system, such as device drivers and network and storage
stacks. Applications communicate with the servers using IPC to access hardware resources.

The \textit{system call} separation layer between user applications and
critical components in monolithic OSes is somewhat arbitrary, chosen for
historical reasons such as rough correspondence to the POSIX standard.
However, more kernel components can be moved into the application itself,
bringing it closer to the hardware.
This concept was initially proposed by Tom Anderson~\cite{LIBRARYANDERSON},
later
implemented in the exokernel model~\cite{EXOKERNEL}, and named
the ``library OS.''
This model is often advocated for performance
reasons and OS
flexibility~\cite{BASCULE}. Subsequent works pointed out the security benefits of the library OS model~\cite{DRAWBRIDGE, GRAPHENE, GRAPHENE_SGX} due to
the strong isolation provided by the reduced interactions between the
application and the privileged layer. Specialized library OSes such as unikernels~\cite{MIRAGE, RUMP,
HERMITCORE, OSV, LING, KYLINX} do not isolate OS components from
applications and are designed
to run just a single application in virtualized environments typical of cloud infrastructures. As unikernels are application-oriented, they may provide
a significant reduction in software stacks compared to
full-fledged OSes.

\subsection{NetBSD, rump kernels, and rumprun}

\textit{Rump kernels} are a concept introduced and popularized by Antti Kantee~\cite{Kantee_flexibleoperating}
and the larger NetBSD~\cite{NETBSD} community.
NetBSD is a well known monolithic OS; along with FreeBSD and OpenBSD, it is
one of
the most popular BSD systems in use today.
NetBSD provides a fairly large collection of device drivers, and
its latest 8.0 version supports state-of-the-art 10GbE networking, NVMe storage, and XHCI/USB 3.0.
Unlike in other OSes, NetBSD's code was largely
redesigned to factor out its device drivers and core
components into \textit{anykernel} components.
As shown in
Figure~\ref{fig:rump}, a special rump kernel glue layer
enables execution of anykernels
outside of the monolithic NetBSD kernel. For example, using POSIX threads
(\textit{pthreads}),
rump kernels can run anykernels in user space on top of existing OSes.
In fact, NetBSD already uses rump kernels to
facilitate its own device driver development and testing process,
which is done much easier in user space.

\rump{}

Rump kernels may serve as a foundation for new OSes, as
in the case of the \textit{rumprun} unikernel.
As shown in Figure~\ref{fig:rump}, systems built from rump kernels
and rumprun are, effectively, library OSes. In contrast to
NetBSD, rumprun-based systems substitute system calls with
ordinary function calls from \textit{libc}.
The original rumprun, unfortunately, lacks SMP support.

Rump kernels were also used to build servers as in the rump file system
server~\cite{RUMPFS}. However, prior solutions simply rely on user-mode
\textit{pthreads}, targeting monolithic OSes. The design
of such servers is very different from a more low-level,
microkernel-based architecture proposed in this paper.

\subsection{Linux-based Library OSes}

We also considered Linux Kernel Library (LKL)~\cite{LKL}, which is based on the Linux source.
However, LKL is not as flexible as rump kernels yet. Additionally, rump kernels
have been upstreamed into the official NetBSD repository,  whereas LKL
is still an unofficial fork of the Linux source branch. rump kernels also support far more applications at the moment.

Unikernel Linux (UKL)~\cite{UKL} uses the unikernel model for Linux by removing
CPU mode transitions. UKL is in an early stage of development; its stated goals are
similar to that of rumprun. However, it is unclear
whether it will eventually match rump kernels' flexibility.

Aside from technical issues, due to
incompatibility, Linux's non-standard GPLv2-only license may
create legal issues\footnote{This is just an opinion of this paper's authors and must not be interpreted as an authoritative legal advice or guidance.} when linking proprietary\footnote{Due to imprecise language,
Linux's syscall exception falls into a gray
legal area when the system call layer is no longer present or clearly defined.} or even
GPLv3+\footnote{Without the syscall exception, GPLv3+ is incompatible with GPLv2 (GPLv2+ only)~\cite{GNULICENSE}. Moreover, most user-space code has moved to GPLv3+.} open-source
applications.
Since licensing terms cannot be adapted easily without explicit permission from \textit{all} numerous Linux contributors, this can create
serious hurdles in the wide
adoption of Linux-based library OS systems.
In contrast, rump kernels and NetBSD use a permissive 2-BSD license.

\subsection{Hypervisors and Microkernels}

\textit{Hypervisors} are typically used to isolate entire guest
OSes in separate \textit{Virtual Machines} (VMs).
In contrast, \textit{microkernels} are typically used to reduce
the \textit{Trusted Computing Base} (TCB) of a single OS and provide component
isolation using \textit{process} address space separation.
Although hypervisors and microkernels have different design goals, they
are inter-related and sometimes compared to each other~\cite{Hand+:hotos2005,Heiser+:osr2006}. Moreover, hypervisors can be used for driver isolation in
VMs~\cite{safe-direct-xen-oasis04}, whereas microkernels
can be used as virtual machine monitors~\cite{Ford+:osdi1996}.

The original uniprocessor rumprun executes as a
paravirtualized (PV)~\cite{Whitaker+:osdi2002} guest OS instance on top
of the Xen hypervisor~\cite{Barham+:sosp03} or
KVM~\cite{kivity07kvm}.
Recently, rumprun was also ported~\cite{SEL4RUMP} to seL4~\cite{SEL4}, a
formally verified microkernel,
where rumprun instances execute as ordinary microkernel processes. The seL4
version of rumprun still lacks SMP support, as it only changes its platform abstraction layer.

\subsection{PCI Passthrough and Input-Output Memory Management Unit (IOMMU)}

VMs can get direct access to physical devices using their hypervisor's PCI passthrough capabilities. 
IOMMU~\cite{INTELIOMMU,AMDIOMMU} hardware support
makes
this process safer by remapping interrupts and I/O addresses used
by devices. When IOMMU is enabled, faulty devices are
unable to inject unassigned interrupts or perform out-of-bound 
DMA operations. Similarly, IOMMU prevents VMs from
specifying out-of-bound addresses, which is especially important
when VM applications are given direct hardware access.
We use both PCI passthrough and IOMMU in the LibrettOS prototype to access physical devices.

\subsection{Single-Root I/O Virtualization (SR-IOV)}

To facilitate device sharing across isolated entities such as
VMs or processes, self-virtualizing
hardware~\cite{RajSchwan:hpdc2007} can be used.
SR-IOV~\cite{SRIOV} is an industry standard that allows splitting one physical PCIe device into logically separate {\it virtual functions} (VFs). Typically, a host OS uses the device's {\it physical function} (PF) to
create and delete VFs, but the PF need not be involved in
the data path of the VF users. SR-IOV is already widely used for high-end
network adapters, where VFs represent logically separate devices
with their own MAC/IP addresses and Tx/Rx hardware buffers.
SR-IOV is also an emerging technology for NVMe storage, where
VFs represent storage partitions.

VFs are currently used by VMs for better network performance,
as each VM gets its own slice of the physical device. Alternatively,
VFs (or even the PF) can be used by the DPDK library~\cite{DPDK}, which
implements lower layers
of the network stack (L2/L3) for certain NIC adapters. High-throughput network
applications use DPDK for direct hardware access.

\subsection{POSIX and Compatibility}

Although high-throughput network applications can benefit
from custom APIs~\cite{IX}, POSIX compatibility remains critical for many
legacy applications. Moreover, modern server applications require
asynchronous I/O beyond POSIX, such as Linux's epoll\_wait(2)
or BSD's kqueue(2) equivalent.
Typically, library OSes implement their own APIs~\cite{IX,ARRAKIS}. Some
approaches~\cite{DRAWBRIDGEPOSIX,GRAPHENE} support POSIX but still run on top of a
monolithic OS with a large TCB. LibrettOS provides full POSIX
and BSD compatibility (except \textit{fork(2)} and \textit{pipe(2)} as discussed in Section~\ref{sec:limit}), while avoiding a large TCB in the underlying
kernel.

\subsection{Summary}

Table~\ref{tbl:comparos} 
summarizes OS design features and
compares LibrettOS against existing
monolithic, library, and multiserver OSes. Since LibrettOS builds on
NetBSD's anykernels, it inherits a very large driver code base.
Applications do not generally need to be modified as the user space
environment is mostly compatible with that of NetBSD, except for 
scenarios described in Section~\ref{sec:limit}. LibrettOS supports two paradigms simultaneously: applications with high performance requirements can directly access
hardware resources as with library OSes (or DPDK~\cite{DPDK} and SPDK~\cite{SPDK}), while other applications can use
servers that manage corresponding resources as with multiserver OSes.
Similar to MINIX 3~\cite{MINIX}, HelenOS~\cite{HELENOS}, VirtuOS~\cite{VIRTUOS}, and Snap~\cite{SNAP}, LibrettOS supports failure recovery
of servers, which is an important feature of the multiserver OS paradigm.
Additionally, LibrettOS's applications can dynamically switch between the
library and multiserver OS modes at run-time.

In Section~\ref{sec:related-work}, we provide a detailed discussion of the past and related work and contrast them with LibrettOS.

\section{Design}

One of the defining features of LibrettOS is its ability to use
the same software stacks and device drivers irrespective of whether an application
accesses hardware devices directly or via servers. In this
section, we describe LibrettOS's architecture, its
network server, and dynamic switching capability.

\subsection{LibrettOS's Architecture}

\librettos{}

Figure~\ref{fig:librettos} presents LibrettOS's architecture. In
this illustration, Application 1 runs in library OS
mode, i.e., it accesses hardware devices such as NVMe and NIC
directly by running corresponding drivers in its own address space.
Since an OS typically runs several applications, hardware resources
need to be shared. In this model, devices are shared using hardware
capabilities such as SR-IOV which splits one NVMe device into
partitions and one NIC into logically separate NICs with dedicated Tx/Rx
buffers. We assume that the NIC firmware does not have fatal
security vulnerabilities -- i.e., it prevents applications from adversely affecting one another.
(VF isolation is already widely used by guest VMs where similar security concerns exist.)

While some high-end NICs expose VFs using SR-IOV, the total number of VFs
is typically limited, and only a small number of applications that
need faster network access will benefit from this hardware capability.
When there are no available VFs for applications, or
their direct exposure to applications is undesirable due to
additional security requirements, we use network servers.
As in library OS mode,
Application 2's data (see Figure~\ref{fig:librettos})  goes through libc sockets and the entire network
stack to generate L2 (data link layer) network frames. However, unlike
Application 1, the network frames are not passed to/from the NIC driver
directly, but rather relayed through a {\it virtual interface} (VIF).
The VIF creates an IPC channel with a network server where frames are
additionally verified and rerouted to/from the NIC driver.

LibrettOS is mostly agnostic to the kind of hypervisor
or microkernel used by rumprun. For easier prototyping while
preserving rich IPC capabilities
and microkernel-like architecture,
we used Xen, which runs rumprun instances in VMs.
Unlike KVM, Xen has a microkernel-style design and does not use any
existing monolithic OS for the hypervisor implementation.
Since rumprun has already been ported to the seL4 microkernel~\cite{SEL4RUMP},
and our core
contributions are mostly orthogonal to the underlying hypervisor, the LibrettOS
design principles can be adopted in seL4 as well.

Scheduling in LibrettOS happens at both the hypervisor and
user application levels. The hypervisor schedules
per-VM \textit{virtual CPUs} (VCPUs) as usual.
Whereas the original rumprun implemented a simple non-preemptive $N:1$
uniprocessor scheduler, we had to replace it entirely to support SMP -- i.e., 
multiple VCPUs per instance.
Each rumprun instance in LibrettOS runs its own $M:N$ scheduler that
executes $M$ threads on top of $N$ VCPUs.

LibrettOS does not have any specific restrictions on the type of
hardware used for direct access; we tested and evaluated both networking
and storage devices. We only implemented a network server to demonstrate indirect access. 
However, storage devices and file systems
can also be shared using an NFS server.
The NFS rumprun instance runs on top of an NVMe device initialized with
the ext3 file system and exports
file system shares accessible through network. In the long term,
an existing rump kernel-based approach~\cite{RUMPFS} for running file systems
in the user space of monolithic OSes can be adopted in the
rumprun environment as well.

LibrettOS does not currently support anything outside of the POSIX/BSD
APIs. Admittedly, POSIX has its downsides -- e.g., copying
overheads which are undesirable for high-performance networking.
Nonetheless, LibrettOS already eliminates the system call
layer, which can open the way
for further optimizations by extending the API in the future.

\subsection{Network Server}

\label{sec:network}

Due to flexibility of rump kernels, their components can be
chosen case by case when building applications. Unlike more traditional OSes,
LibrettOS runs the TCP/IP stack
directly in the application address space.
As we discuss below, this does not compromise security
and reliability. In fact, compared to typical multiserver and monolithic OSes, where buggy network stacks affect all applications, LibrettOS's network stack issues are localized to one application.

The network server IPC channel consists of Tx/Rx ring buffers which
relay L2 network
frames in \textit{mbufs} through a {\it virtual network interface} (VIF)
as shown in Figure~\ref{fig:netdom}.
We use a lock-free ring buffer implementation from~\cite{RINGBUFFER}.
When relaying mbufs, the receiving side,
if inactive, must be notified using a corresponding notification mechanism.
For the Xen hypervisor, this is achieved using a virtual interrupt (VIRQ).

Since a stream of packets is generally transmitted without delays, the
number of such notifications is typically small.
The sending side detects that
the receiving
side is inactive by reading a corresponding shared (read-only) status variable
denoted as \textit{threads}, which indicates how many
threads are processing requests. When this variable is zero, a VIRQ is sent
to wake a worker thread up.

The network server side has a per-application \textit{handler}
which forwards mbufs to the NIC driver.
Frames do not need to be translated,
as applications and servers are built from the same rump kernel source base.

\netdom{}

\begin{figure}[ht!]
\centering
\includegraphics[width=.6\columnwidth]{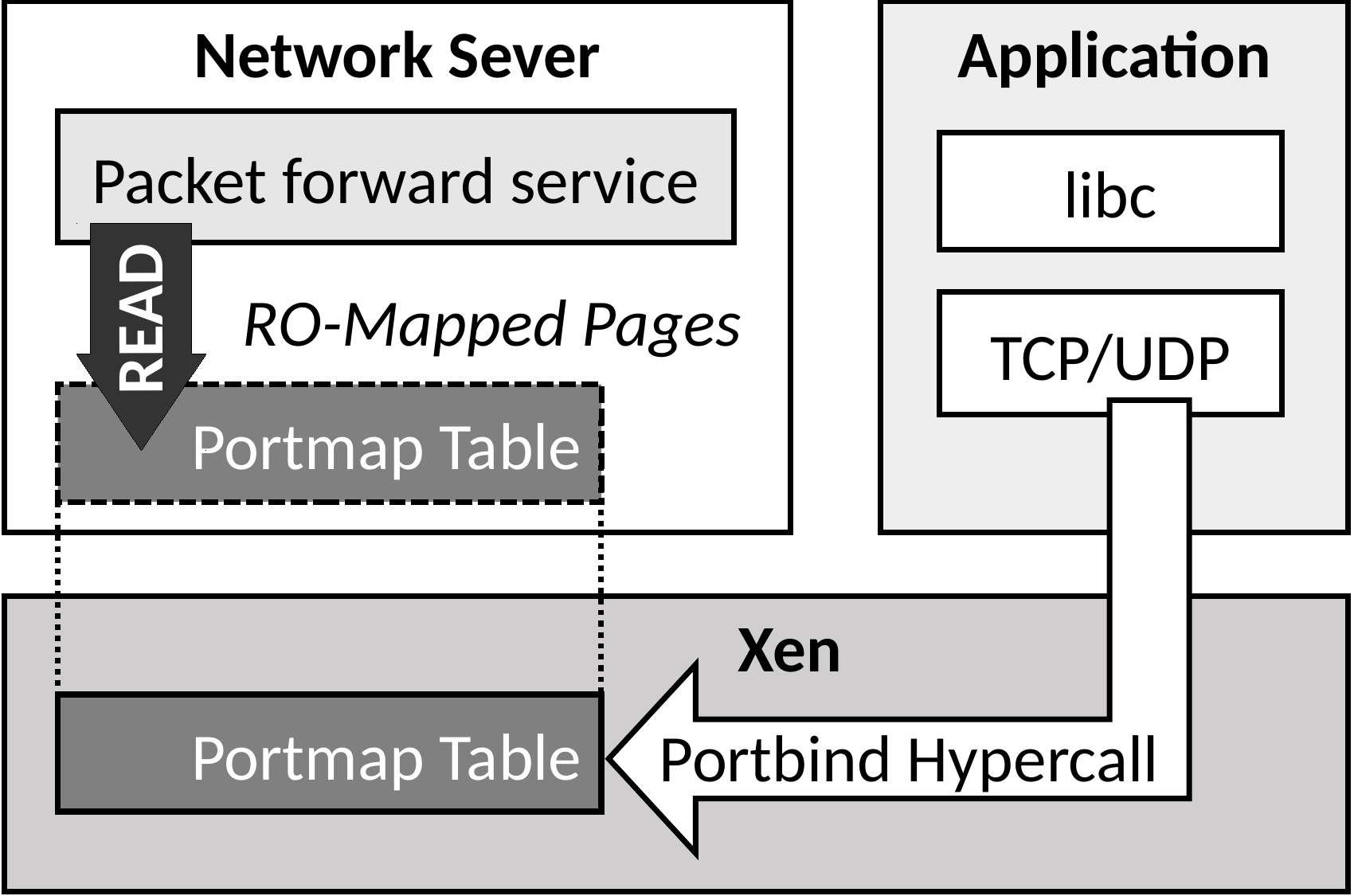}
\caption{Portmap table.}
\label{fig:portmap_table}
\end{figure}

An important advantage of this design is that an application
can recover from any network server failures, as TCP state
is preserved inside the application. The network server operates at the very
bottom of the network stack and simply routes inbound
and outbound mbufs. However, routing across different applications requires
some knowledge of the higher-level protocols.
We support per-application routing based on TCP/UDP ports.
When applications
request socket bindings, the network server allocates corresponding ports
(both static and dynamic)
by issuing a special portbind hypercall as shown in Figure~\ref{fig:portmap_table}
which updates a per-server 64K port-VM (portmap) table in the hypervisor.
We store the portmap in the hypervisor to avoid its loss
when the network server is restarted due to failures.
While this mechanism is specific to the network server, the added code
to the hypervisor is very small and is
justified for omnipresent high-speed networking. Alternatively, the portmap table can be moved to a
dedicated VM.

Since packet routing requires frequent reading of the portmap entries,
we read-only map the portmap pages into the network server address space, so
that the network server can directly access these entries.

The network server does not
trust applications for security and reliability reasons. We enforce
extra metadata verification while
manipulating ring buffers. Since applications may still transmit
incorrect data (e.g., spoof MAC/IP source addresses), either automatic
hardware spoofing detection must be enabled, or address
verification must be enforced by verifying MAC/IP addresses
in outbound frames. Similarly, ports need to be verified by the server.

Although, unlike POSIX, invalid packets may be
generated by (malicious) applications at any point, they can only appear
in local (per-application) ring buffers. When the network server copies
packets to the global medium, it verifies
MAC, IP, and port metadata which must already be valid for a given
application. Races are impossible as prior users of the same MAC, IP,
and port must have already transmitted all packets to the
global medium and closed their connections.

We also considered an alternative solution based on the netfront/netback model~\cite{Barham+:sosp03}
used by Xen. The netback driver typically runs
in an OS instance with
a physical device (Driver Domain).
The netfront driver is used by guest OSes as a virtual NIC to establish
network connectivity through the netback driver. Rumprun already implements a
very simple netfront driver so that it can access a virtual NIC when running
on top of the Xen hypervisor. However, it lacks the netback driver that would
be required for the server.
This model is also fundamentally different from our approach:
transmitted
data must traverse additional OS layers on the network server side
and all traffic needs to go through network address
translation (NAT) or bridging, as each application (VM instance) gets its own
IP address. While the Xen approach is suitable when running
guest OSes, it introduces unnecessary layers of indirection
and overheads as we show in Section~\ref{exp:netfrontres}.

\subsection{Dynamic Mode Switch}

LibrettOS's use of the same code base for both direct and server modes
is crucial for implementing the dynamic mode switch.
We implement this feature for networking, which is more or less stateless.
As NVMe's  emerging support for SR-IOV becomes publicly available, our 
technique can also be considered for storage devices. However, storage
devices are more stateful and additional challenges need to be overcome.

\begin{figure}[ht!]
\centering
\includegraphics[width=.9\columnwidth]{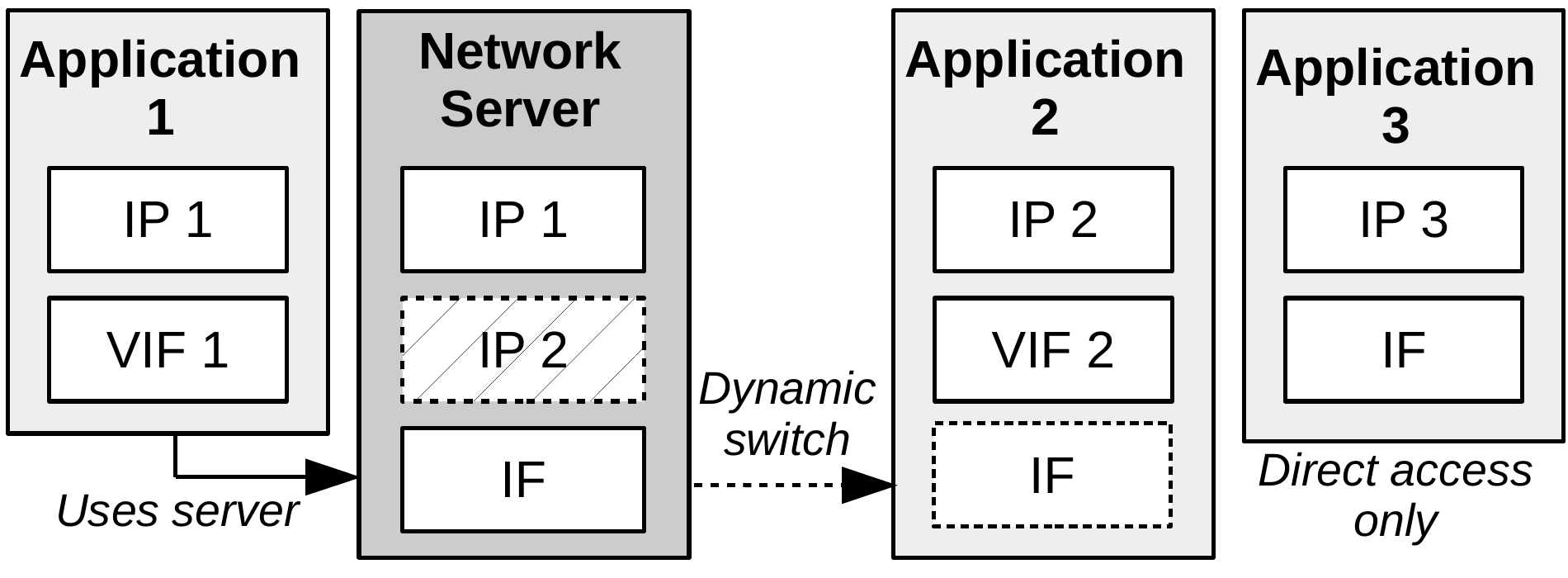}
\caption{Dynamic switching mechanism.}
\label{fig:dynamic}
\end{figure}

Figure~\ref{fig:dynamic} shows LibrettOS's  dynamic switching mechanism for the network server. LibrettOS is the only OS that supports this mechanism.
Moreover, applications do not need to be modified to benefit from this feature.
As we further discuss in Section~\ref{exp:evalsect}, dynamic switching  does not incur large overhead, and switching
latency is small (56-69$ms$).

The biggest challenge for dynamic switching is
that application migration must be completely transparent. When using
direct access (VF), we need a dedicated IP address
for each NIC interface. On the other hand, our network server shares
the same IP address across all applications. A complicating factor is
that once connections are established, IP addresses used by applications
cannot be changed easily.

To allow dynamic switching, we
leverage NetBSD's support for multiple IPs per an interface (IF).
Unlike hardware Rx/Tx queues,
IPs are merely logical resources; they are practically unlimited when using NAT.
A pool of available IPs is given to a server, and addresses are recycled as connections are closed.
When launching an application that needs dynamic switching
(Application~2), the network server allocates an IP address that
will be used by this application. Applications use virtual interfaces (VIF) which
are initially connected to
the network server (i.e., similar to Application~1). The network
server adds the IP to the list of
configured addresses for its IF (PF or VF).

As I/O load increases, an application can be decoupled from the network
server to use direct access for better throughput and CPU utilization.
The NIC interface used by the network server deactivates application's
IP (which remains \textit{unchanged} throughout the lifetime of an application). The application then configures an available
NIC interface (typically, VF) with its dedicated IP.
Changing IP on an interface triggers ARP invalidation.
A following ARP probe detects a new MAC address for the interface-assigned IP.
These events are masked from applications, as
they typically use higher-level protocols such as TCP/IP. No pause-and-drain
is needed, and on-the-fly requests
are simply ignored because TCP automatically resends packets without ACKs.

When running in this mode, all traffic from the NIC interface (direct access)
is redirected through VIF, as all application socket bindings belong to this VIF.
This special VIF is very thin: \textit{mbufs} are redirected to/from IF without
copying, i.e., avoiding extra overhead. When the load decreases,
applications can return to the original state by performing the above
steps in the reverse order.
If applications always use direct access (Application~3), VIFs are not needed, i.e., all traffic goes directly through IF.

Dynamic switching helps conserve SR-IOV resources, i.e., it enables more effective use of available hardware resources across different applications. Policies for switching (e.g., for managing resources, security, etc.) are beyond the scope of this paper, as our focus is on the switching mechanism. In our evaluation, users manually trigger dynamic switching.

\section{Implementation}

While NetBSD code and its rump kernel layer are SMP-aware
(except rump kernel bugs that we fixed), none of the
existing rumprun versions (Xen PV, KVM, or seL4) support SMP.

\subsection{Effort}

We redesigned low-level parts of rumprun (SMP BIOS initialization, CPU initialization,
interrupt handling) as well as higher-level components such as the
CPU scheduler.
Our implementation avoids
coarse-grained locks on performance-critical paths.
Additionally, the original rumprun was only
intended for paravirtualized Xen. Thus, we added  support for SMP-safe, 
Xen's HVM-mode \textit{event channels} (VIRQs) and
\textit{grant tables} (shared
memory pages) when implementing our network server.
We implemented the Xen HVM interrupt callback mechanism by binding it to the
per-CPU interrupt vectors as it is implemented in Linux and FreeBSD.
The HVM mode uses hardware-assisted virtualization and is more preferable
nowadays for security reasons~\cite{MELTDOWN}. Moreover, it has better
IOMMU support when using device PCI passthrough.
We also fixed race conditions in the rump layer, and added NVMe and
10GbE ixgbe driver glue code which was previously unavailable for
rumprun. Additionally, we ported a patch that adds support
for the EVFILT\_USER extension in kqueue(2), which is required by some
applications and supported by FreeBSD but is, for some reason,
still unsupported by NetBSD.

Since rumprun does not currently support preemption, we designed
a non-preemptive, lock-free M:N scheduler with global and per-VCPU queues. A lock-free
scheduler has an
advantage that, it is not adversely affected by preemption which still
takes place in the hypervisor. (When the hypervisor de-schedules a VCPU holding a lock, other VCPUs cannot get the
same lock and make further progress.) Lock-free data structures
for OS schedulers were already
advocated in~\cite{OSV}.

In Table~\ref{tbl:impl}, 
we present the effort that was required
to implement our current LibrettOS prototype. Most changes
pertain to rumprun and the network server implementation, whereas drivers
and applications did not require substantial changes.

\begin{table}
\begin{center}
\begin{tabular}{ l l r }
  \toprule
	\textbf{Component} & \textbf{Description} & \textbf{LOC} \\
  \midrule
	SMP support & SMP BIOS init/trampoline, & 2092 \\
	& M:N scheduler, interrupts &  \\
	HVM support & Xen grant tables and VIRQ & 1240 \\
	PV clock & SMP-safe Xen/KVM clock & 148 \\
	Network server & mbuf routing, & 1869  \\
        & dynamic switching & \\
	Xen & Network server registrar & 516 \\
	NetBSD, network & Network server forwarder & 51 \\
	NetBSD, kqueue & EVFILT\_USER support & 303 \\
	NetBSD, rump & Race conditions in SMP & 45  \\
	NetBSD, glue & New drivers (NVMe, ixgbe) & 234 \\ 
    NFS, rpcbind, & Porting from NetBSD, & 154 \\
    $ $ and mountd & avoiding fork(2) & \\
  \midrule
	\textbf{Total:} & & \textbf{6652} \\
  \bottomrule
\end{tabular}
\end{center}
\caption{Added or modified code.}
\label{tbl:impl}
\end{table}

\subsection{Limitations}

\label{sec:limit}

In our current prototype, applications cannot create processes
through \textit{fork(2)}, as its implementation would require complex
inter-VM interactions which Xen is not designed for.
We consider such an implementation to be realistic for
microkernels
such as seL4.
Each application, however, can already support virtually
unlimited number of threads.
Moreover, typical applications that rely on fork~\cite{APACHE}
can be configured to use the multi-threading mode instead.

We did not implement POSIX's IPCs such as
\textit{pipe(2)}, but Xen-based mechanisms can be used with no restrictions.

As \textit{fork(2)} is
not currently supported, we did not consider sharing socket file descriptors or \textit{listen(2)}/\textit{accept(2)} handoff across processes.
This may create additional challenges in the future.
We believe that the problem is tractable, but requires modification
of the network stack.
Also, Rx/Tx ring buffers will have to be created per connection
rather than per process to allow sharing per-socket/connection buffers
across processes while not violating process isolation requirements.

As previously mentioned, we did not develop policies for dynamic switching as our focus was on mechanism.
Section~\ref{exp:evalsect} uses a simple policy wherein a user triggers an event.

\section{Evaluation}

\label{exp:evalsect}

We evaluate LibrettOS using a set of micro- and macro-benchmarks and compare
it against NetBSD and Linux. Both of these baselines are relevant since
LibrettOS is directly based on the NetBSD code base, and Linux is a popular
server OS. LibrettOS typically has better performance, especially when
using direct mode.
Occasionally, Linux demonstrates better performance since it currently has
more optimizations than NetBSD in its drivers and network stack,
especially for 10GbE+, as we further discuss in Section~\ref{sec:performnetbsd}.

In the evaluation, we only consider standard POSIX compatible applications,
as compatibility is one of our important design goals. Since kernel-bypass
libraries and typical library OSes~\cite{DPDK,SPDK,IX} are not compatible
with POSIX, they cannot be directly compared. Moreover, kernel-bypass
libraries require special TCP stacks~\cite{MTCP} and
high engineering effort to adopt existing
applications~\cite{Zhang:2019:IDY:3317550.3321422}.
Likewise, comparison against existing multiserver designs~\cite{MINIX,SNAP}
is challenging due to very limited application and device driver support.

\begin{table}
\begin{center}
\begin{tabular}{ l l }
  \toprule
Processor & 2 x Intel Xeon Silver 4114, 2.20GHz \\
Number of cores & 10 per processor, per NUMA node \\
HyperThreading & OFF (2 per core) \\
TurboBoost & OFF \\
L1/L2 cache & 64 KB / 1024 KB per core \\
L3 cache & 14080 KB \\
Main Memory & 96 GB \\
Network & Intel x520-2 10GbE (82599ES) \\
Storage & Intel DC P3700 NVMe 400 GB \\
\bottomrule
\end{tabular}
\end{center}
\caption{Experimental setup.}
\label{tbl:system}
\end{table}

Table~\ref{tbl:system} shows our experimental setup.
We run Xen 4.10.1, and Ubuntu 17.10 with
Linux 4.13 as Xen's Dom0 (for system initialization only).
We use the same version of Linux as our baseline.
Finally, we use NetBSD 8.0 with the
NET\_MPSAFE feature\footnote{NET\_MPSAFE, recently introduced in NetBSD, reduces global SMP locks in the network stack. By and large, the network stack is the only major place where global locks still remain. The ongoing effort will eliminate them.} enabled for better network scalability~\cite{NETBSD8}.
LibrettOS uses the same NetBSD code base, also with NET\_MPSAFE enabled.
We set the maximum transmission unit (MTU) to 9000 in all experiments
to follow general recommendations~\cite{10gbe:linux2009} for the optimal 10GbE
performance.
We measure each data point 10 times and present the average; the relative standard deviation is mostly less than 1\%.

LibrettOS not only outperforms NetBSD, but also outperforms
Linux in certain tests. This is despite the fact that NetBSD 
can be slower than Linux. LibrettOS's advantage comes
from its superior multiserver-library OS design and
better resource management (compared to NetBSD).

\subsection{NetBSD and Linux performance}

\label{sec:performnetbsd}

Since neither the original rumprun nor other unikernels support
10GbE+ or NVMe physical devices,
one of our goals is to show how this cutting-edge hardware
works in LibrettOS.

Although NetBSD, Linux, and rumprun's 1GbE are known to be on
par with each other~\cite{SEL4RUMP},
10GbE+ networking puts more pressure on
the system, and we found occasional performance
gaps between NetBSD and Linux. However, they are mostly due to specific
technical limitations, most
of which are already being addressed by the NetBSD community.

One contributing factor is the
network buffer size. NetBSD uses \textit{mbuf} chains; each \textit{mbuf}
is currently limited to 512 bytes in x86-64. In contrast, Linux has larger (frame-sized) \textit{sk\_buff} primitives.
Since 512-byte units are too small for 10GbE+ jumbo frames, we increased the
\textit{mbuf} size to 4K; this improved performance by $\approx$10\% on
macrobenchmarks. \textit{mbuf} cannot be increased beyond the 4K page size
easily without bigger changes to the
network stack, but this is likely to be addressed by the NetBSD community sooner
or later.

Another contributing factor is the use of global SMP locks by NetBSD. Since
NET\_MPSAFE was introduced very recently, and the effort to remove the remaining
coarse-grained locks is still ongoing, Linux's network stack
and network drivers may still have superior performance in some tests
which use high-end 10GbE+ networking.

We also found that the adaptive interrupt moderation option in NetBSD's
ixgbe driver adversely affected performance in some tests such as NetPIPE
(e.g., NetBSD was only able to reach half of the throughput).
Thus, we disabled it.

\subsection{Sysbench/CPU}

To verify our new scheduler and SMP support, we ran
the Sysbench/CPU~\cite{web:sysbench} multi-threaded test which measures the
elapsed time
for finding prime numbers. Overall,
LibrettOS, Linux, and NetBSD show roughly similar performance.

\subsection{NetPIPE}

\label{exp:netfrontres}

\begin{figure}[ht!]
\centering
\subfigure[Throughput]{
\includegraphics[width=.88\columnwidth]{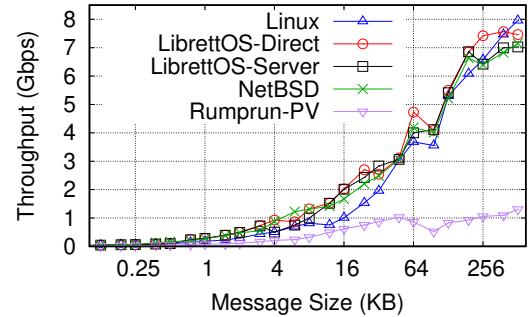}
\label{fig:expernetpipe}
}
\\
\vspace{-7pt}
\subfigure[CPU utilization (1 CPU is 100\%)]{
\includegraphics[width=.88\columnwidth]{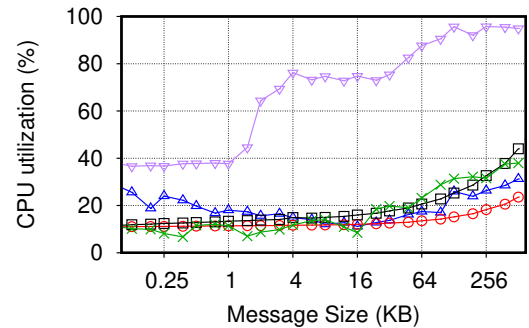}
\label{fig:expernetpipecpu}
}
\caption{NetPIPE (log2 scale).}
\end{figure}

To measure overall network performance, we use NetPIPE, a popular ping pong benchmark. It exchanges a fixed size
message between
servers to measure the latency and bandwidth of a single flow. 
We evaluate Linux, LibrettOS (direct access mode), LibrettOS (using our network
server), NetBSD, and the original rumprun with the netfront driver.
On the client
machine, Linux is running for all the cases.
We configure NetPIPE to
send 10,000 messages from 64 bytes to 512KB.
Figure~\ref{fig:expernetpipe} shows the throughput for different message sizes. 
LibrettOS (direct access) achieves 7.4Gbps with 256KB. It also 
has a latency of 282$\mu$s for 256KB messages.
LibrettOS (network server) achieves 6.4Gbps with 256KB, and
the latency of 256KB messages is 326$\mu$s. Linux
achieves 6.5Gbps with 256KB messages and the latency of 318.9$\mu$s.
NetBSD reaches 6.4Gbps in this test with the latency of
326$\mu$s.
The original rumprun with the netfront driver reaches only 1.0Gbps with
256KB messages and has a large latency of 1,983$\mu$s.

Overall, all systems except the original rumprun have comparable performance.
With smaller messages ($<$ 256KB), LibrettOS-Direct can be a bit faster than Linux or NetBSD.
We attribute this improvement to the fact that for smaller messages,
the cost of system calls is non-negligible~\cite{Soares+:osdi10}.
In LibrettOS, all system calls are replaced with regular function calls.
Even LibrettOS-Server avoids IPC costs, as
the server and application VMs typically run on different CPU cores.

\begin{figure}[ht!]
\centering
\includegraphics[width=\columnwidth]{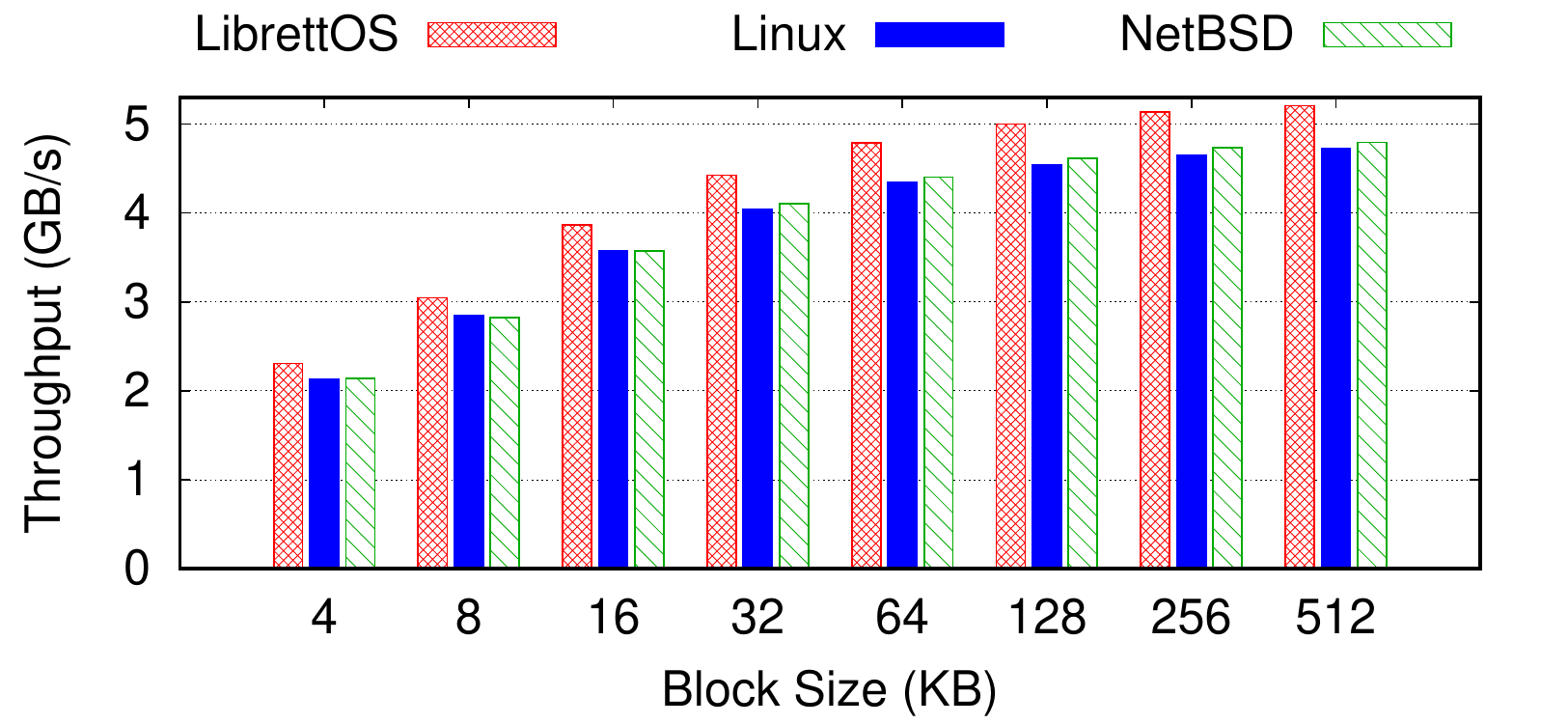}
\caption{NFS server.}
\label{fig:expernfs}
\end{figure}

LibrettOS-Direct is more efficient not just on  throughput but, more importantly, on CPU utilization (Figure~\ref{fig:expernetpipecpu}). The saved CPU resources can be used in myriad ways -- e.g., running more applications, obtaining higher overall performance.
For LibrettOS-Server, there is a 1.8x gap in overall CPU consumption due
to an extra layer where packets are transferred from an application
to the server.
The same argument applies to other experiments in this section -- e.g., dynamic switching.

\subsection {NFS Server}

To evaluate NVMe storage, we run an NFS server macrobenchmark by
using Sysbench/FileIO~\cite{web:sysbench} from the client.
We measure mixed file I/O composed of page cache 
and storage I/O because users benefit from the page cache in general. 
We mount an NVMe partition initialized with the ext3 file system and export
it through the NFS server. For LibrettOS, we use direct NIC and NVMe access.
We mount the provided NFS share on the client side and run the Sysbench test
with different block sizes (single-threaded).
Figure~\ref{fig:expernfs} shows that
for all block sizes, LibrettOS outperforms both NetBSD and Linux.
LibrettOS is 9\% consistently faster than both of them. This is because, 
LibrettOS avoids the system call layer which is known to cause a
non-negligible performance overhead~\cite{Soares+:osdi10}.

\subsection {Nginx HTTP Server}

\label{exp:nginxhttp}

\begin{figure}[ht!]
\centering
\includegraphics[width=\columnwidth]{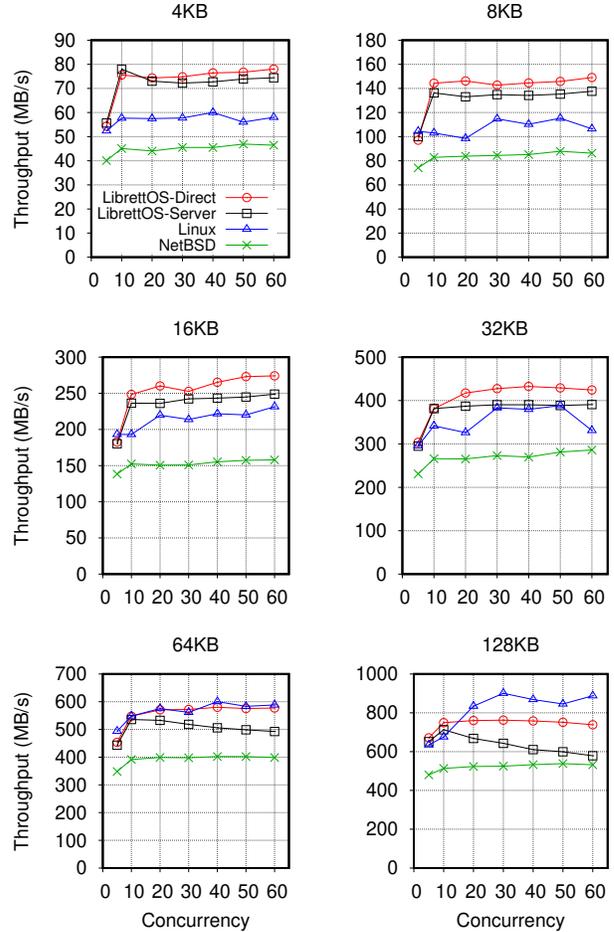}
\caption{Nginx HTTP server.}
\label{fig:expernginx}
\end{figure}

To run a network-bound macrobenchmark, we choose Nginx 1.8.0~\cite{NGINX}, a
popular web server. We use the Apache Benchmark~\cite{APACHEBENCH} to generate
server requests from the client side.
The benchmark is set to send 10,000 requests with various levels of
concurrency on the client side. (In this benchmark, concurrency
is the number of concurrent requests.)
Figure~\ref{fig:expernginx} shows the throughput with
different concurrency levels and file sizes.
LibrettOS (direct access and server) is always faster than NetBSD.
Except very large blocks (128KB), LibrettOS-Direct also outperforms Linux.
Particularly, for 8KB/50,
LibrettOS (direct access) is 66\% faster than NetBSD and 27\% faster than
Linux.
Even LibrettOS-Server outperforms Linux in many cases
due to its optimized, fast IPC mechanism.
We note that LibrettOS is able to outperform Linux even though
the original NetBSD is slower than Linux in this test.
We generally attribute this improvement to the fact that LibrettOS removes the
system call layer. Furthermore, compared to NetBSD, LibrettOS improves
scheduling and resource utilization.
For very large blocks (128KB), NetBSD's network stack current limitations
outweigh other gains.

\begin{figure}[ht!]
\centering
\includegraphics[width=\columnwidth]{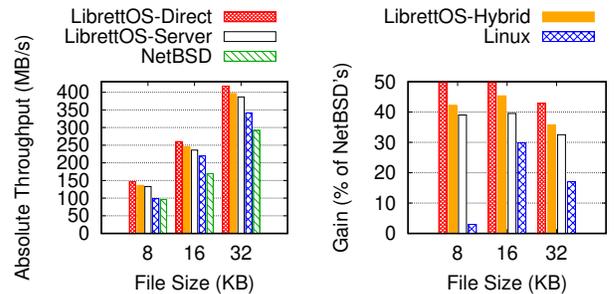}
\caption{Dynamic switch (Nginx).}
\label{fig:dynamicadapt}
\end{figure}

We also evaluated dynamic switching when running Nginx. In Figure~\ref{fig:dynamicadapt},
LibrettOS-Hybrid represents a test
with runtime switching from the server to direct mode and back to the server mode (the concurrency level is 20).
Overall, Nginx spends 50\% of time in the server mode and 50\% in the direct mode.
LibrettOS-Hybrid's average throughput shows benefits of using this mode compared to LibrettOS-Server, Linux,
and NetBSD. The relative gain (Figure~\ref{fig:dynamicadapt}) over Linux, let alone NetBSD, is quite
significant in this test.
We also measured the cost of dynamic switching to be
56$ms$ from the server to direct mode, and 69$ms$ in the opposite direction.

\subsection{Memcached}

\begin{figure}[ht!]
\centering
\includegraphics[width=\columnwidth]{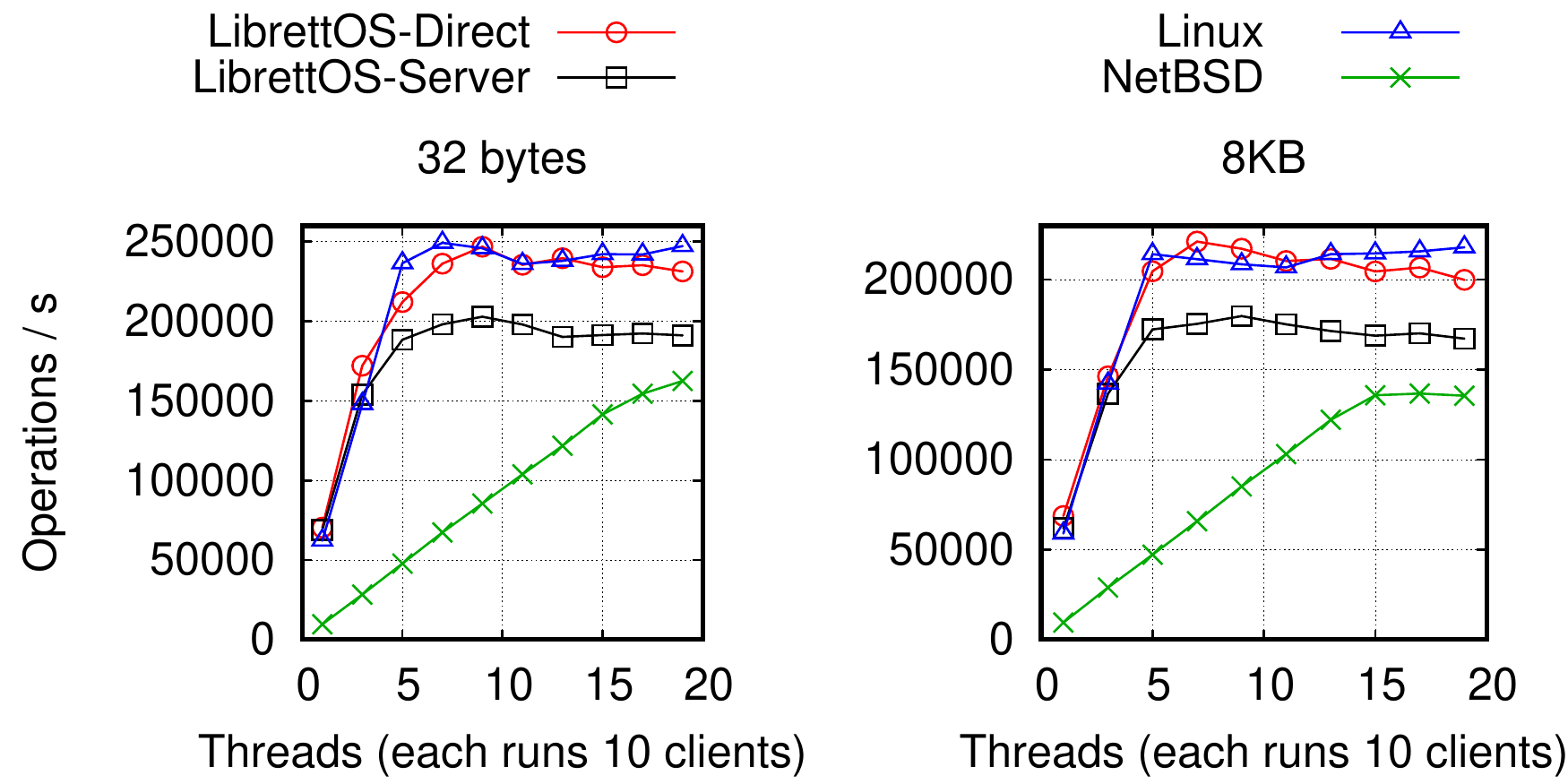}
\caption{Memcached (distributed object system).}
\label{fig:expermemcached}
\end{figure}

We also tested LibrettOS's network performance with memcached~\cite{MEMCACHED}, a
well-known distributed memory caching system, which caches objects in RAM to
speed up database accesses.
We use memcached 1.4.33 on the server side. On the client side,
we run memtier\_benchmark 1.2.15~\cite{MEMTIER} which acts as a load generator.
The benchmark performs SET and GET operations using the \textit{memcache\_binary} protocol. We use the default ratio
of operations 1:10 to ensure a typical, read-dominated workload. In Figure~\ref{fig:expermemcached}, we present the
number of operations per second for a small and large block sizes
and using a different number of threads (1-20). Each
thread runs 10 clients, and each client performs 100,000 operations.

LibrettOS-Direct and Linux show similar overall performance. LibrettOS-Server
reveals a runtime overhead since it has to run a server
in a separate VM. Finally, NetBSD shows the worst performance as
in the previous test.

\subsection {Redis}

\begin{figure}[ht!]
\centering
\includegraphics[width=\columnwidth]{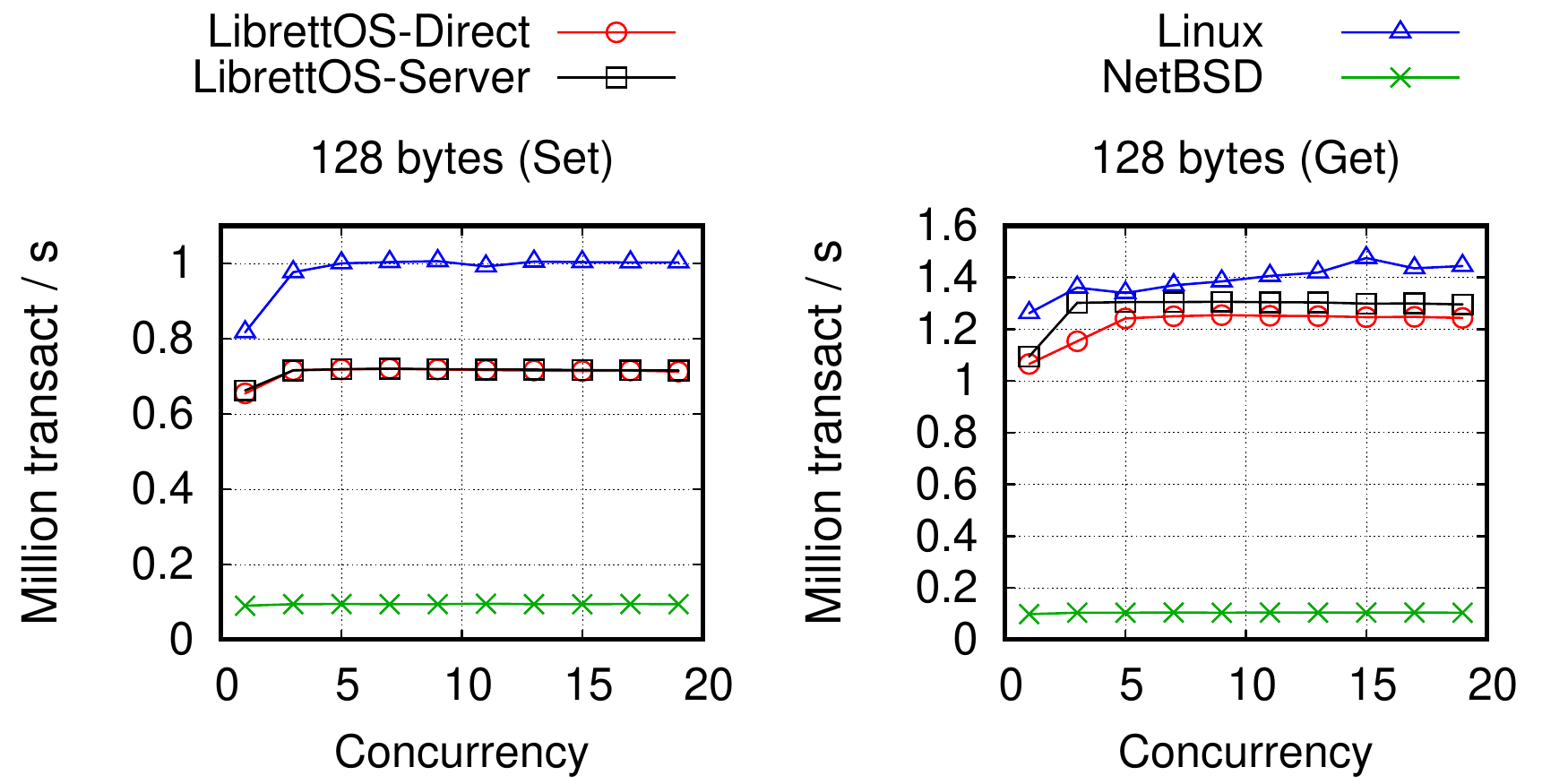}
\caption{Redis key-value store.}
\label{fig:experredis}
\end{figure}

Finally, we ran Redis, an in-memory key-value store used by many cloud applications~\cite{REDIS}.
We use Redis-server 4.0.9 and measure the throughput of 1,000,000 SET/GET operations. 
In the pipeline mode, the Redis benchmark sends requests without waiting for
responses which improves performance. We set the pipeline size to 1000.
We measure the performance of Redis for various number of concurrent
connections (1-20).
We show results for the SET/GET operation (128 bytes); trends for other sizes
are similar.
LibrettOS's GET throughput (Figure~\ref{fig:experredis}) is close to Linux,
but the SET throughput is smaller. NetBSD reveals very poor performance
in this test, much slower than that of LibrettOS and Linux. We suspect
that the current Redis implementation is suboptimal for certain BSD
systems, which results in worse performance that we did not
observe in other tests.

\subsection {Failure Recovery and Software Upgrades}

LibrettOS can recover from any faults occurring in its servers,
including memory access violations, deadlocks, and  interrupt handling routine failures. Our model assumes that failures are localized to one server, i.e.,
do not indirectly propagate elsewhere.
Applications that use our network server do not necessarily fail: the portmap
state is kept in the hypervisor, and if recovery happens quickly, TCP can still simply resend packets without timing out (applications keep their TCP states).
Additionally, when upgrading the server, an artificial fault can be injected
so that an application can use a newer version of the server without restarting the application itself.

\begin{figure}[ht!]
\centering
\subfigure[Single application (Nginx)]{
\includegraphics[width=.47\columnwidth]{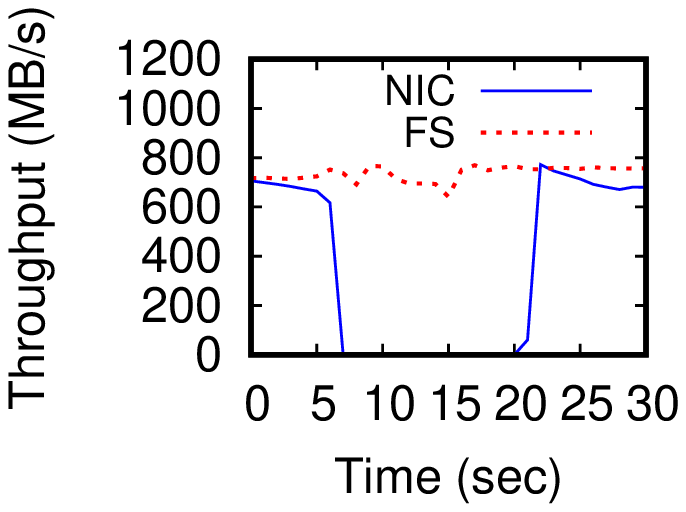}
\label{fig:experfailure}
}
\subfigure[Multiple applications]{
\includegraphics[width=.47\columnwidth]{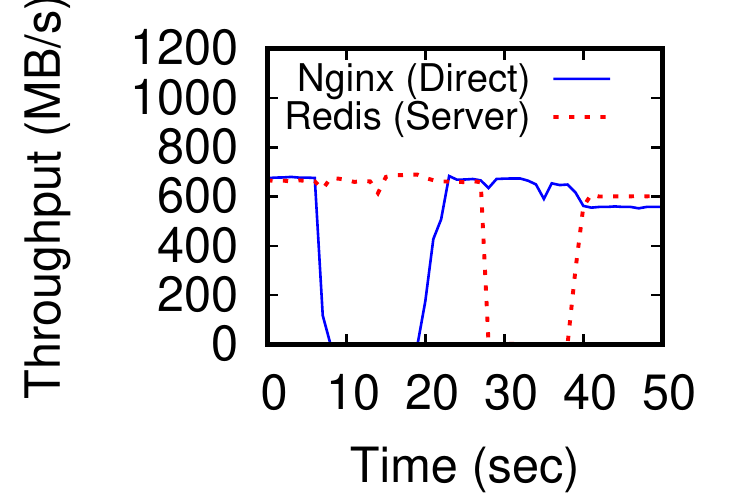}
\label{fig:experfailuremulti}
}
\caption{Failure recovery and transparent upgrades.}
\end{figure}

To demonstrate LibrettOS's 
failure recovery (or transparent server upgrades), we designed two experiments.
In the first experiment, an application has two
jobs: one uses networking (Nginx)
and the other one accesses a file system. When LibrettOS's network server
fails, the Nginx server is unable to talk to the NIC hardware, and we report
outage that is observable on the client side. This failure, nonetheless, does
not affect the file system job. In Figure~\ref{fig:experfailure}, we
show the corresponding network and file system transfer speeds. After rebooting the network server, Nginx can again reach the network, and clients continue their data transfers.
In the other test, Figure~\ref{fig:experfailuremulti}, we run two applications.
Nginx uses direct access while Redis uses our network server. We show that a
failure of Nginx does not affect Redis. Then we also show that a network server
failure does not impact Nginx.

\section{Related Work}\label{sec:related-work}

LibrettOS's design is at the intersection of three OS research topics:
system component isolation for security, fault-tole\-rance, and transparent software upgrades (multiserver OS);
app\-lication-specialized OS components for performance, security, and software bloat reduction (library OS);
and kernel bypassing with direct hardware access to applications for performance.

Isolation of system software components in separate address spaces is the key principle of the microkernel model~\cite{NUCLEUS, MINIX, L4, MACH,
HELENOS, QNX, SINGULARITY}. Compared to the classical monolithic OS design,
microkernels provide stronger security and reliability guarantees~\cite{SEL4,
MINIX_FAULT_ISOLATION, MINIX_ASLR}. With microkernels, only essential
functionalities (scheduling, memory management, IPC) are implemented within the kernel.
L4~\cite{L4} is a family of microkernels, which is known to be
used for various purposes. A notable member of this family
is seL4~\cite{SEL4}, a formally verified microkernel.
Multiserver OSes such as MINIX 3~\cite{MINIX},
GNU Hurd~\cite{Bushnell:hurd1996}, Mach-US~\cite{Stevenson+:usenix1995}, and 
SawMill~\cite{Gefflaut+:eurosigops2000} are a specialization of
microkernels where OS components (e.g., network and storage stacks, device
drivers) run in separate user processes known as ``servers.''

Decomposition and isolation have proved to be beneficial for various layers of
system software in virtualized environments:
hypervisor~\cite{DECONSTRUCTING_XEN}, management toolstack~\cite{XOAR}, and
guest kernel~\cite{VIRTUOS}. Security and reliability are particularly
important in virtualized environments due to their multi-tenancy
characteristics and due to the increased reliance on cloud computing for today's
workloads. In the desktop domain, Qubes OS~\cite{QUBEOS} leverages Xen to run
applications in separate VMs and provides strong isolation for local
desktop environments. LibrettOS also uses Xen to play the role of the
microkernel. However, the LibrettOS design is independent
of the microkernel used and can be applied as-is to other microkernels
supporting virtualization, with potentially smaller code bases~\cite{NOVA, L4RE,SEL4}.

The network stack is a crucial OS service, and its reliability has
been the subject of multiple studies based on microkernels/multiserver
OSes using state replication~\cite{CURIOS},
partitioning~\cite{NEAT}, and  checkpointing~\cite{GIUFFRIDA_OS_CRASH}.
Several user-space network stacks such as mTCP~\cite{MTCP}, MegaPipe~\cite{MEGAPIPE},
and OpenOnload~\cite{OPENONLOAD} have been designed to bypass the OS 
and avoid various related overheads. While it is not their primary objective,
such user-space network stacks do provide a moderate degree of reliability:
since they do not reside in the kernel, a fault in the network stack will not
affect the applications that are not using it.

In its multiserver mode, LibrettOS obtains the
security and reliability benefits of microkernels by decoupling the application
from system components such as drivers that are
particularly prone to faults~\cite{DRIVERS_BUGS1, DRIVERS_BUGS2} and
vulnerabilities~\cite{DRIVERS_VULN}.

IX~\cite{IX} and Arrakis~\cite{ARRAKIS} bypass traditional
OS layers to improve network performance compared to commodity OSes.
IX implements its own custom API while using the DPDK library~\cite{DPDK}
to access NICs. Arrakis supports POSIX but builds on top of
the Barrelfish OS~\cite{BARRELFISH}, for which device driver support is limited.
LibrettOS gains similar performance benefits when using direct hardware access
(library OS) mode, while reducing code development and maintenance costs as it reuses a very large base of drivers and software from a popular OS, NetBSD.
Additionally, in contrast to LibrettOS, these works do not consider
recoverability of OS servers.

Certain aspects of the multiserver design can be employed using existing
monolithic OSes. VirtuOS~\cite{VIRTUOS} is a fault-tole\-rant multiserver design based on Linux that provides
strong isolation of the kernel components by running them in virtualized service domains on top of Xen. Snap~\cite{SNAP} implements a network server
in Linux to improve performance and simplify system upgrades. However, Snap uses its private
(non-TCP) protocol which cannot be easily integrated into existent applications.
Moreover, Snap is limited to networking and requires re-implementing all network device drivers.

On-demand virtualization~\cite{Kooburat:2011:BBW:1991596.1991602} implements
full OS migration, but does not specifically target SR-IOV devices.
In contrast, LibrettOS allows dynamic device migrations.

To optimize I/O in VMs, researchers have proposed the sidecore
approach~\cite{Gavrilovska07high-performancehypervisor,Kumar2007RearchitectingVF}
which avoids VM exits and offloads the I/O work to sidecores. Since this approach is wasteful when I/O activity reduces,
Kuperman et al.~\cite{Kuperman:2016:PRI:2872362.2872378}
proposed to consolidate sidecores from different machines onto a single server.

Exokernel~\cite{EXOKERNEL} was among the first to propose compiling
OS components as libraries and link them to applications. Nemesis~\cite{NEMESIS} implemented a library OS with an extremely lightweight kernel.
Drawbridge~\cite{DRAWBRIDGE}, Graphene~\cite{GRAPHENE}, and Graphene-SGX~\cite{GRAPHENE_SGX}
are more recent works that leverage the security benefits of library OSes
due to the increased isolation resulting from the lesser degree of interaction
between applications and the privileged layer.
Bascule~\cite{BASCULE} demonstrated OS-independent extensions for library OSes.
EbbRT~\cite{EBBRT} proposed a framework for building per-app\-lication library OSes for performance.

Library OSes specialized for the cloud --
unikernels~\cite{MIRAGE, RUMP, HERMITCORE, OSV, LING, KYLINX} --  have also emerged in
recent years. They are dedicated to running in the cloud and are a practical
implementation of the library OS model where the hypervisor plays the role
of the exokernel. In its library OS mode, LibrettOS builds upon and reaps the
benefits of this model. Additionally, LibrettOS gains
the same benefits of multiserver OSes such as failure recovery. LibrettOS is
also the first SMP design that runs unikernels with unmodified drivers
that belong to the hardened NetBSD driver code base.

Red Hat has recently initiated an effort, which is in its early stage, to
create a Linux unikernel (UKL)~\cite{UKL}. UKL promises to resolve the
problem of maintaining a large base of drivers and specialized I/O stacks
(e.g., DPDK~\cite{DPDK}, SPDK~\cite{SPDK}, and mTCP~\cite{MTCP}).
UKL's long-term goal of replacing DPDK and SPDK by using
unikernelized applications is already achieved in the LibrettOS design,
which also encapsulates a more powerful multiserver/library OS design.

\section{Conclusion}

We presented LibrettOS, an OS design that unites two fundamental
models, the multiserver OS and the library OS, to reap their combined benefits. In
essence, LibrettOS uses a microkernel to run core OS components as servers
for better isolation, reliability, and recoverability. For selected applications, LibrettOS acts as a
library OS such that these applications are given direct access to I/O devices
to improve performance. We also support dynamic runtime switching between
the two modes. In LibrettOS's unique design, not only most OS code stays
identical as an application switches between different modes of operation,
 more importantly, the application does not have to be modified either.

We built a prototype of the LibrettOS design based on rump kernels. LibrettOS is compatible both
in terms of BSD and POSIX user-space applications, and also in terms of the large driver code base of the NetBSD kernel.
Our prototype implementation necessitated a
significant redesign of the rumprun environment which executes rump kernels, in
particular, the implementation of SMP support.
To demonstrate the multiserver paradigm, we implemented a
network server based on direct L2 frame forwarding.
We evaluated this prototype and
demonstrated successful recovery from OS component failures (and
transparent server upgrades) as well as
increased I/O performance compared to NetBSD and Linux, especially when using direct access.

Finally, although kernel-bypass advantages are widely discussed in the literature,
unlike prior works~\cite{IX,SNAP,ARRAKIS}, LibrettOS is the first to achieve
greater performance with ordinary POSIX applications and out-of-the-box device drivers.

\section*{Availability}

LibrettOS is available at: \url{https://ssrg-vt.github.io/librettos}

\section*{Acknowledgements}

We would like to thank the anonymous reviewers and our shepherd Dilma Da Silva for their insightful comments and suggestions, which helped greatly improve this paper. We also thank Pierre Olivier for his comments and suggestions.

This research is based upon work supported by the Office of the Director of National Intelligence (ODNI), Intelligence Advanced Research Projects Activity (IARPA). The views and conclusions contained herein are those of the authors and should not be interpreted as necessarily representing the official policies or endorsements, either expressed or implied, of the ODNI, IARPA, or the U.S. Government. The U.S. Government is authorized to reproduce and distribute reprints for Governmental purposes notwithstanding any copyright annotation thereon.

This research is also based upon work supported by the Office of Naval Research (ONR) under grants N00014-16-1-2104, N00014-16-1-2711, and N00014-18-1-2022.

\bibliography{os}

\end{document}